# Anomalous charge transport in dodecaborides RB$_{12}$ (R- Ho, Er, Tm, Lu)

N. Sluchanko[1*], L. Bogomolov[1,2], V. Glushkov[1,2], S. Demishev[1,2], M. Ignatov[1,2], N. Samarin[1],

D. Sluchanko[1], A. Levchenko[3], N. Shitsevalova[3], K. Flachbart[4]

[1]*A.M.Prokhorov General Physics Institute of RAS, 38, Vavilov str., Moscow, 119991, Russia*

[2]*Moscow Institute of Physics and Technology, 9, Institutskii per., Dolgoprudny,*

*Moscow region, 141700, Russia*

[3]*Institute for Problems of Materials Science, 03680, Kiev, Ukraine*

[4]*Centre of Low Temperature Physics, IEP SAS and IPS FS UPJS, SK-04001Košice, Slovakia*

High precision measurements of Hall $R_H(T)$ and Seebeck $S(T)$ coefficients have been carried out for the first time on single crystals of rare earth dodecaborides RB$_{12}$ (R - Ho, Er, Tm, Lu) at temperatures 1.8 - 300K. Low temperature anomalies associated with antiferromagnetic phase transitions in HoB$_{12}$, ErB$_{12}$ and TmB$_{12}$ compounds have been detected on the temperature dependencies of $R_H(T)$ and $S(T)$. The estimated values of charge carriers' mobility allowed us to conclude about the appreciable influence of spin fluctuations on the charge transport in these compounds with B$_{12}$ atomic clusters.

**1.** In the family of rare earth and transition metal borides a growing interest of researchers is attracted to the study of compounds with framework structural units - boron nanoclusters B$_6$ and B$_{12}$. Depending on the cubic crystal structure parameters and the filling of internal *4f–* and valence *5d,6s*–shells of rare earth ions, these compounds reveal dielectric ($SmB_6$ and $YbB_{12}$) or metallic ($R^{3+}B_6$, $R^{3+}B_{12}$) types of conductivity at low temperatures with superconducting ($YB_6$, $ZrB_{12}$, $LuB_{12}$ *et al.*) and various magnetic (para-, dia-, ferro- and antiferromagnetic) ground states. Similar to the intensively studied perovskites as well as molybdates and tungstates, these "framework structures" are characterized by a cooperative motion of connected polyhedral boron units, and seem to be perspective model objects for studying the nature of magnetic interactions and the peculiarities of ground state formation in 3D systems with nanoclusters and localized magnetic moments due to their simple crystal structure (*bcc* for hexaborides $RB_6$ and *fcc* for dodecaborides $RB_{12}$ ). In this respect one of the most important and fundamental problems is connected with the identification of the interplay between phonon spectra and electronic band structures of these crystals, their magnetic properties, and the features of charge transport in these strongly correlated metallic systems.

[*] E-mail: nes@lt.gpi.ru

As far as we know, there are no trustworthy data of Hall and Seebeck coefficients behaviour in rare earth dodecaborides at present time. The current status of experimental studies is given mainly by considerable obstacles which result from (i) particular complications of the growth of $RB_{12}$ single crystals due to high melting temperatures ($T_m \geq 2300$ K) and (ii) due to high conductivities, which impede precision measurements of transport coefficients in these metallic compounds.

2. The aim of the experimental study was to investigate in details the behaviour of Hall ($R_H$) and Seebeck (*S*) coefficients as well as the resistivity ($\rho$) in paramagnetic and antiferromagnetic phases of rare earth dodecaborides $RB_{12}$ (R - Ho, Er, Tm, Lu) at helium and intermediate temperatures (1.8 – 300 K). Single crystals of $RB_{12}$ grown by the crucible-less inductive zone melting in argon gas atmosphere [1, 2] have been used for measurements. The Hall coefficient $R_H(H,T)$ has been measured by the sample rotation technique with the step-by-step fixation of the sample's position in steady magnetic field $H \sim 3\text{-}4$ kOe using the experimental setup described in [3]. The original four-probe technique with the variation of temperature gradient on the sample [4] was applied to study the thermopower. All the measurements have been performed with temperature gradient $\nabla T$ in (110) plane and sample current $\boldsymbol{I} \parallel <110> \perp \boldsymbol{H}$.

**3.** The angular dependencies of Hall resistivity $\rho_H(\varphi)$ measured for some $RB_{12}$ compounds are presented in fig.1. It was found that in the whole temperature interval corresponding to the paramagnetic state of rare-earth dodecaborides the $\rho_H(\varphi)$ dependence can be well described by the cosine law $\rho_H(\varphi) \sim \rho_{H1} \cos\varphi$. Results of straightforward fitting have been used to estimate the absolute value of Hall coefficient from the simple relation $\rho_{H1} = R_H d$ ($d$ − the sample's thickness). The calculated values of Hall coefficient measured in the paramagnetic phase of $RB_{12}$ (R- Ho, Er, Tm, Lu) turn out to be rather small ($R_H(T) \sim 2.8 \div 4.5 \cdot 10^{-4}$ cm$^3$/Coul, see also fig. 2) with a relative changes of $R_H(T)/R_H(300K)$ not exceeding the maximal value of ~30% estimated for $HoB_{12}$. However, the onset of antiferromagnetic (AF) ordering below Nèel temperature $T_N$ is accompanied by considerable distortions of the $\rho_H(\varphi)$ dependencies that is shown by Hall effect data obtained for $ErB_{12}$ (fig.1). The appearance of even (second $\rho_H(\varphi) \sim \rho_{H2}\cos2\varphi$ and forth $\rho_H(\varphi) \sim \rho_{H4}\cos4\varphi$) harmonics, which are similar to those earlier observed in antiferromagnetic state of $CeAl_2$ [3], seriously complicates the analysis of $R_H(T)$ dependence at $T < T_N$ (fig.1) The discussion of even Hall effect harmonics occurring in these strongly correlated electron systems is outside the scope of this paper and will be presented in details elsewhere.

The temperature dependencies of Hall coefficient $R_H(T)$ obtained in this study for $RB_{12}$ compounds (R - Ho, Er, Tm, Lu) are presented in fig. 2. As seen from data of fig. 2, the behaviour of $R_H(T)$ for some magnetic ($HoB_{12}$, $ErB_{12}$) dodecaborides in general is roughly similar to that of non-magnetic ($LuB_{12}$) compound at intermediate temperatures (20 K ≤ $T$ ≤ 300 K). A wide maximum of $R_H(T)$ observed at temperatures 20 ÷ 50K is evolved to the minimum detected in the range of 100÷300K. In the case of TmB$_{12}$, the observed feature is transformed in the sequence of three extremas (see the lowest curve in fig. 2). However, the temperature independent behaviour of the Hall coefficient, which is observed in non-magnetic metal LuB$_{12}$ at low temperatures $T < 20K$ (fig. 2), is not specific for the magnetic dodecaborides under investigation. Particularly, distinct maxima of $R_H(T)$ are observed at $T = T_{max} \sim 10\text{-}12$ K for HoB$_{12}$ and TmB$_{12}$. Additionally, the pronounced anomalies of Hall coefficient associated with AF transitions for all magnetic compounds RB$_{12}$ can be clearly seen on the temperature dependences of $R_H(T)$ in the vicinity of Neel temperatures $T_N$. The estimated values of $T_N$, which are equal to 7.4 K, 6.7 K and 3.3 K for HoB$_{12}$, ErB$_{12}$ and TmB$_{12}$, respectively, agree very well to the results of magnetic and thermodynamic measurements of RB$_{12}$ [5-7]. Note also that the drastic increase of Hall coefficient observed for the first time in AF phase of ErB$_{12}$ (fig. 2) requires further investigation of Hall effect in the magnetic phase of this compound.

The temperature dependencies of Seebeck coefficient $S(T)$ as measured for magnetic and non-magnetic compounds RB$_{12}$ (R - Ho, Er, Tm, Lu) also demonstrate the consistent behaviour (fig. 3). A pronounced feature of Seebeck coefficient at intermediate temperatures (100 – 300 K) is observed for all dodecaborides under investigation, the amplitude of the negative minimum of $S(T)$ being the largest one for non-magnetic lutetium dodecaboride (fig. 3). When lowering temperature, the thermopower of all magnetic dodecaborides changes the sign. In the case of HoB$_{12}$ and TmB$_{12}$, a wide positive maximum of relatively small amplitude $S \leq 0.5$ µV/K is detected on $S(T)$ curves in the temperature interval of 10÷20 K (inset of fig. 3). For ErB$_{12}$, the sign inversion and the maximum of $S(T)$ are observed at temperatures 10.5 K and 8 K, respectively (fig. 3). The transition into AF state in RB$_{12}$ is accompanied by the appearance of an additional positive contribution to Seebeck coefficient (inset of fig. 3). The values of Nèel temperature for ErB$_{12}$ and HoB$_{12}$ found from the approximation of $S(T)$ curves in AF-phase to temperature axis (see inset of fig. 3) are in good consistence to the $T_N$ values estimated from Hall effect (fig. 2), and magnetic [5] and thermal [6] measurements. In the case of TmB$_{12}$, the vicinity of $T_N \approx 3.3$ K is characterized by a very small amplitude maximum on the $S(T)$ curve. The value of $T_N^S$ determined by the mentioned approximation procedure is lower than $T_N$ - $T_N^S$(TmB$_{12}$) $\approx$ 2.7 K (inset of fig. 3).

To receive additional information about the transport properties of RB$_{12}$ single crystals the temperature dependencies of resistivity $\rho(T)$ have been measured for all the samples under investigation. The results of the $\rho(T)$ study (see fig. 4a) allowed us to estimate the values of $T_N$, which agree with previous results [5-7] within the experimental accuracy.

**4.** When analysing the temperature behaviour of transport coefficients in RB$_{12}$ compounds, it should be taken into account that trivalent rare earth dodecaborides in the paramagnetic phase are metals having a rather complicated Fermi surface (FS) consisting of two parts. The first part is multiply connected in the <111> directions and topologically similar to the FS of copper, the second one forms pancake-like electron surfaces centred at the X point [8]. Band structure calculations and quantum oscillations studies performed on LuB$_{12}$ [8] allowed to conclude that the conduction band of LuB$_{12}$ is mainly formed by the *5d*-states of rare earth ions hybridised with boron *2p*-states. According to the current conception trivalent rare earth compounds RB$_{12}$ are one-electron metals [5-8], the change of *4f*–state configuration in the sequence of Ho - Tm, Lu having only weak influence on the conduction band structure.

In this respect it is reasonable to estimate the behaviour of Hall mobility in RB$_{12}$ using the common expression for one type of charge carriers $\mu_H(T)=R_H(T)/\rho(T)$. The data of fig. 4b presented for temperatures 1.8 – 300 K demonstrate that the charge carriers mobility reaches its maximum value (~2500 cm$^2$/(V·s)) for the non-magnetic metal LuB$_{12}$ with fully filled *4f*-states of rare earth ions (*4f$^n$*, $n = 14$). Additionally, the change of *4f*-shell configuration in the sequence HoB$_{12}$ - LuB$_{12}$ starting from HoB$_{12}$ (*4f$^n$*, $n = 10$) induces a significant decrease of $\mu_H$ which depends monotonously on the filling number *n* in the range $10 < n < 13$ (fig. 4b). Note that the de-Gennes factor $(g-1)^2J(J+1)$ ($J$ - full magnetic moment of the *4f*-shell), which characterizes the magnetic scattering of charge carriers, is also reduced dramatically with the increase $n$ in the range $10 < n < 13$. It is of special interest to accent here that the intermediate valence state of rare earth ions in dodecaborides ($\upsilon(Yb) \approx 2.9$ [9]) realized in $YbB_{12}$ ($4f^n$, $n = 13$) results to the unusual semiconducting ground state [10].

To explain the anomalous behaviour of transport coefficients in the sequence of $HoB_{12}$ - $YbB_{12}$, it is necessary to propose that the decrease of charge carriers mobility with the increase of $n$ in the range $n = 10$-$13$ seems to be connected with the enhancement of spin (and charge- in case of $YbB_{12}$) fluctuations in these compounds. As a result, the natural explanation can be also provided for (i) the decrease of the $S(T)$ absolute value, which is observed at intermediate temperatures 50-300 K (fig. 3) and accompanies the mobility lowering between $HoB_{12}$ and $TmB_{12}$ (fig. 4b), as well as for (ii) the reduction of Néel temperature from 7.4 K ($HoB_{12}$) down to 3.3 K ($TmB_{12}$) with the increase of $n$, which is, however, much faster than expected for these magnetic dodecaborides (see [2, 5]). Besides, within the framework of the spin fluctuation approach the similar behaviour of $R_H(T)$ curves detected for various compounds $RB_{12}$ in the temperature interval 50÷300 K (fig. 2) as well as that of $S(T)$ dependencies for magnetic and non-magnetic dodecaborides in this range of temperatures (fig. 3) could be associated with the absence of noticeable changes in the structure of *5d-2p*-states forming the conduction band in these rare earth dodecaborides. However, to check the approach proposed in the present work in detail, a comprehensive study of transport, magnetic and thermodynamic characteristics should be performed for $RB_{12}$ in a wide range of temperatures and magnetic fields.

This work was performed in framework of RFBR 04-02-16721 and INTAS 03-51-3036 projects, as well as under the financial support of the RAS Program "Strongly Correlated Electrons in Semiconductors, Metals, Superconductors and Magnetic Materials" and Russian Science Support Foundation. Acknowledged is also the support of Slovak Scientific Grant

\* e-mail: nes@lt.gpi.ru

---

**Figure captions**

**Fig. 1.** Angular dependencies of Hall resistivity $\rho_H(\varphi)$ for $LuB_{12}$ (left panel) and $ErB_{12}$ (right panel). Solid lines present the results of harmonic analysis (see text).

**Fig. 2.** Temperature dependencies of Hall coefficient $R_H(T)$ for $RB_{12}$ compounds (R - Ho, Er, Tm, Lu).

**Fig. 3.** Temperature dependencies of Seebeck coefficient $S(T)$ for $RB_{12}$ compounds (R - Ho, Er, Tm, Lu). $T^*$ corresponds to the peculiarity observed on $S(T)$ curves at intermediate temperatures. The temperature interval in the vicinity of magnetic transitions is presented in the inset in larger scale.

**Fig. 4.** Temperature dependencies of (a) resistivity $\rho(T)$ and (b) Hall mobility $\mu_H(T) = R_H(T)/\rho(T)$ for $RB_{12}$ compounds (R - Ho, Er, Tm, Lu). The numbers on $\mu_H(T)$ curves correspond to the exponents $\alpha$ of the power law dependencies $\mu_H(T) \sim T^{-\alpha}$ fitted in the temperature range 50 - 300 K.

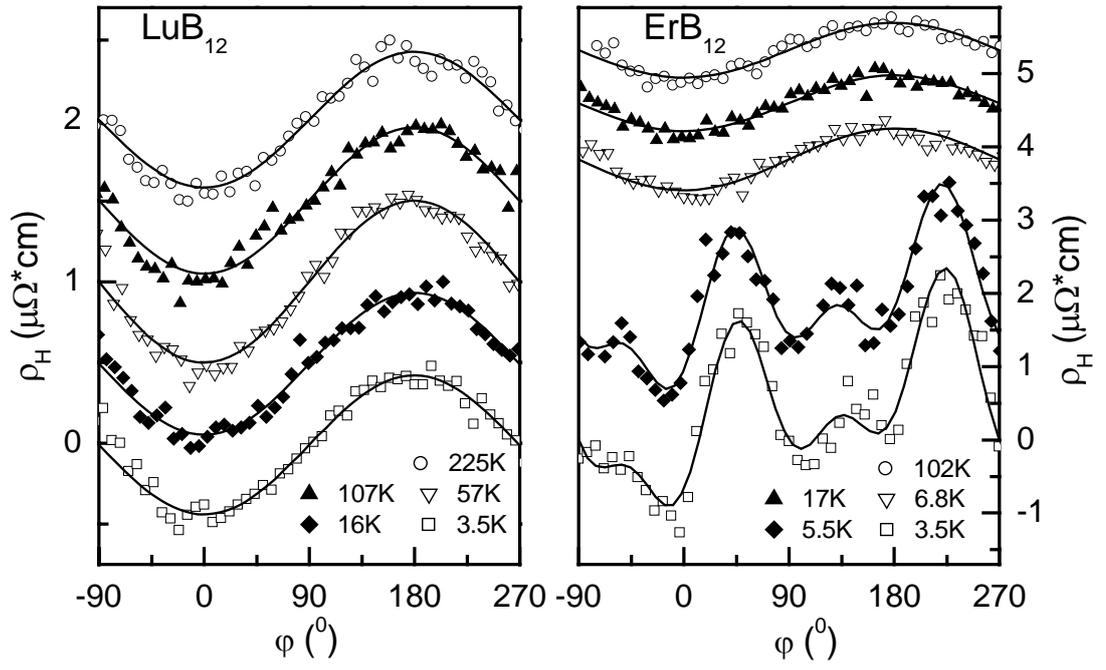

Fig.1

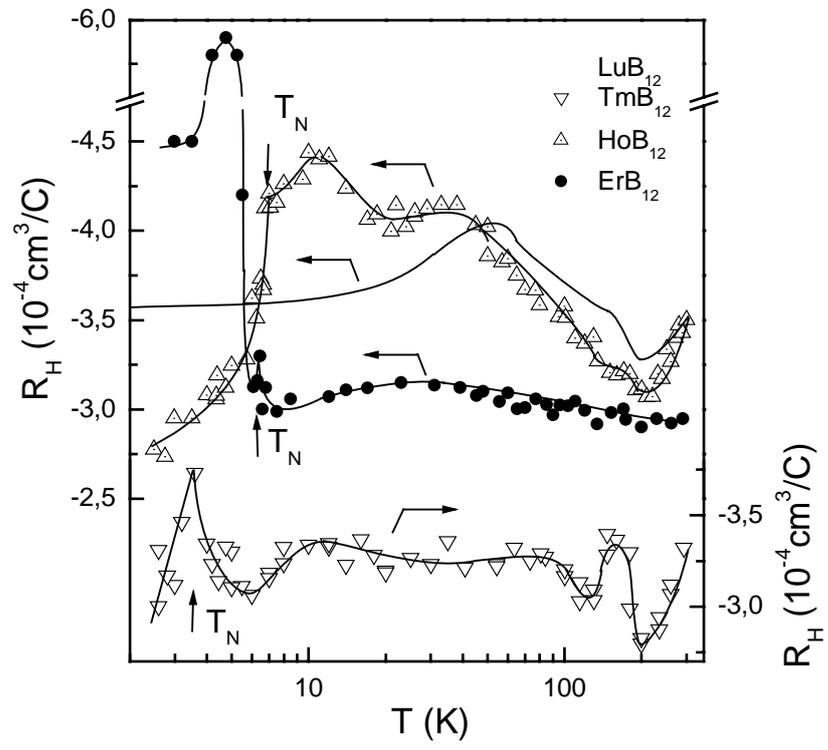

Fig.2

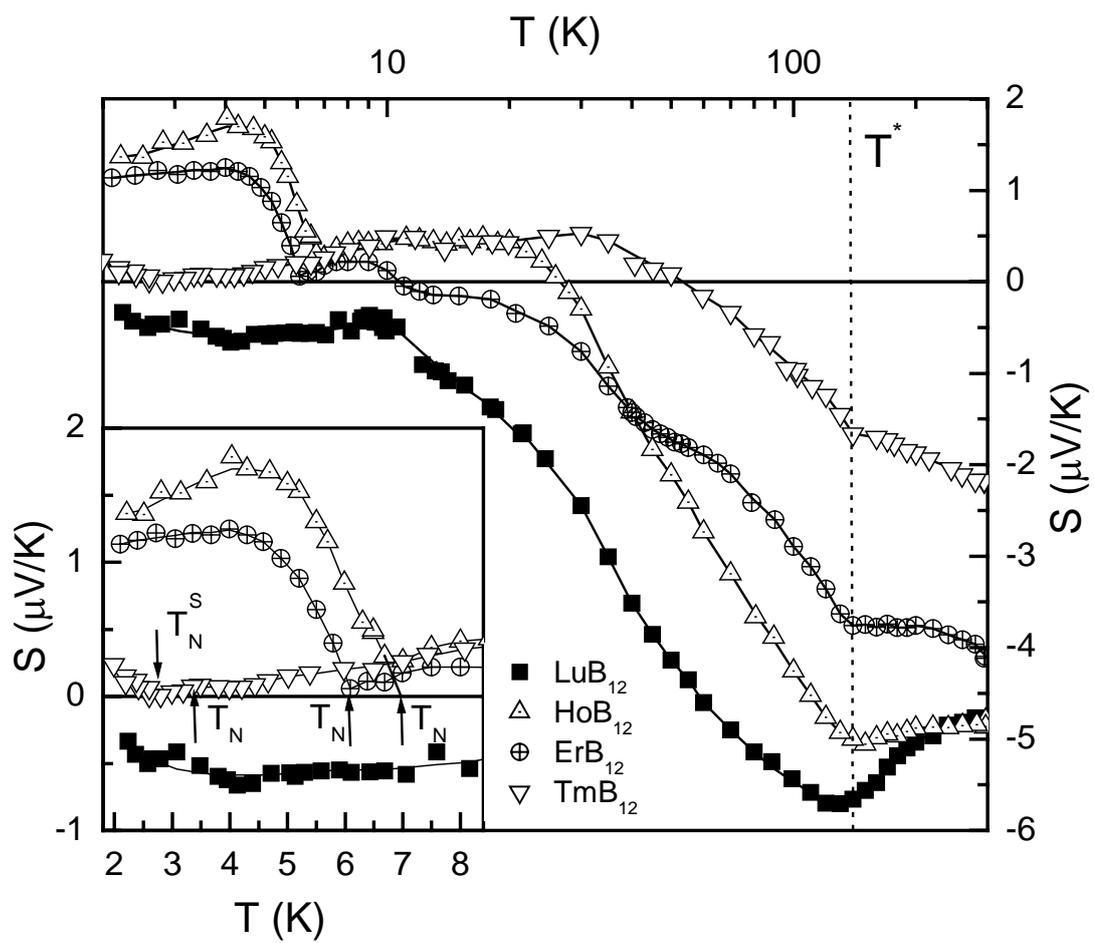

Fig.3

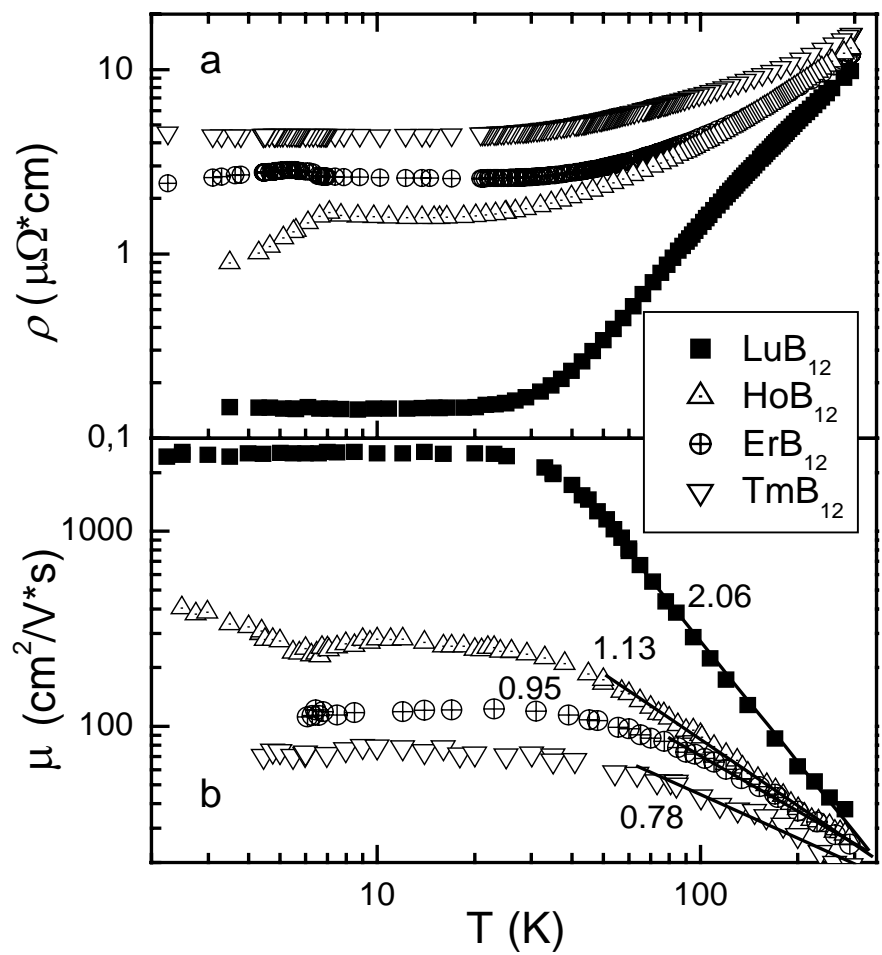

Fig.4